\documentclass[referee,epsfig]{mn2e}

\title[The GRB early optical flashes from internal shocks]{The GRB early optical
flashes from internal shocks: application to GRB990123, GRB041219a and GRB060111b}

\author[D.M. Wei]
{D. M. Wei$^{1,2,3}$\\
$^1$ Purple Mountain Observatory, Chinese Academy of Science,Nanjing 210008, China.\\
$^2$ National Astronomical Observatories,Chinese Academy of Sciences, Beijing, 100012, China.\\
$^3$ Joint Center for Particle Nuclear Physics and Cosmology of
Purple Mountain Observatory - Nanjing University, Nanjing
210008,China.}

\begin{document}

\maketitle

\begin{abstract}
With the successful launch of the Swift Gamma-Ray Burst Explorer, it
is widely expected that the prompt optical flashes like GRB990123
would be easily detected. However, the observations show that for a
number of GRBs no early optical flash has been detected, which
indicates that the reverse shock emission must be suppressed. Here
we explore the possibility that the optical flash may arise from the
internal shock emission. For GRB990123 and GRB060111b, although
their optical emission are not correlated with the gamma-ray
emission, we propose here that their optical and gamma-ray emission
may arise from different internal shocks (which can be formed by
collision of different shells), and find that, under certain
circumstances, the optical flashes of GRB990123 and GRB060111b can
well be explained by the internal shock model. For GRB041219a, the
prompt optical emission was correlated with the gamma-ray emission,
which can also be explained by the internal shock model if we assume
the optical emission was the low energy extension of the gamma-ray
emission, and we find its redshift is about $z\sim 0.2$. As for
GRB050904, we have shown in previous paper that the optical flash
was produced by synchrotron radiation and the X-ray flare was
produced by the synchrotron-self-Compton mechanism. Therefore we
conclude that the early optical flashes of GRBs can usually arise
from the internal shock emission. Meanwhile in our model since the
shells producing the optical flashes would be easily disrupted by
other shells, so we suggest that the bright optical flash should not
be common in GRBs. In addition, we also discussed the
synchrotron-self-Compton emission in the internal shock model, and
find that for different values of parameters, there would be several
kinds of high energy emission (at $\sim 100$ KeV, $\sim 10$ MeV or
GeV) accompanying the optical flash. For a burst like GRB990123, a
GeV flare with fluence about $10^{-8}$ erg cm$^{-2}$ s$^{-1}$ is
expected, which might be detected by the GLAST satellite.
\end{abstract}


\section{Introduction}
Gamma-ray bursts (GRBs) are the most luminous explosions in the
universe, but the origin of their emission is still unclear. With
the successful launch of the Swift Gamma-ray Burst Explorer, great
progress has been made in the study of the early afterglow of GRBs.
The observation of the early afterglow would offer the possibility
to clarify the question whether the early emission is from the
internal shock or from the reverse shock.

The early optical flash of GRB990123 is widely believed to be
produced by the reverse shock emission (Sari \& Piran 1999), and it
is widespread expected that the prompt optical flash like GRB990123
would be easily detected by Swift. However up to now there are only
a few gamma-ray bursts whose prompt optical flashes have been
detected contemporaneous with the high energy emission. For
GRB990123 and recently discovered GRB060111b, their optical flashes
were uncorrelated with the prompt gamma-ray emission, which suggests
that the optical emission and gamma-ray emission should have
different origin (Akerlof et al. 1999; Kulkarni et al. 1999; Klotz
et al. 2006). For GRB041219a, its optical flash was correlated with
the gamma-ray emission (Vestrand et al. 2005; Blake et al. 2005),
and for GRB050904, a very bright optical flare was temporal
coincident with an X-ray flare (Bo\"{e}r et al. 2006), which implies
that for these two GRBs there should be a common origin for the
optical and high energy emission.

If the reverse shock emission is strong, as expected from the
optical flash of GRB990123, then it is naturally to expect that the
early optical emission would be easily detected in the Swift era.
However, the observations show that for a number of GRBs no early
optical emission has been detected, which implies that the reverse
shock emission should be significantly suppressed (Roming et al.
2005). Another possibility is that the optical flash may be produced
by the internal shock emission. M\'esz\'aros \& Rees (1999) have
shown that the internal shock model can well explain the temporal
behavior of the optical flash of GRB990123. As for GRB050904, Wei et
al. (2006) have shown that, within the context of internal shock
model, the optical flash was produced by synchrotron radiation, and
the X-ray flare was produced by the synchrotron-self-Compton
mechanism.

In this paper we will discuss the prompt optical emission and high
energy emission based on the internal shock model. For GRB990123 and
GRB060111b, their optical emission are uncorrelated with the
gamma-ray emission, one possibility is that their optical emission
are from the reverse shock, and also there is another possibility:
the optical and gamma-ray emission are from the different internal
shocks. It is well known that in the internal shock model, the
central engine emits a lot of shells with different Lorentz factors,
so it is natural that there would be many internal shocks formed by
collision of different shells, for example, the gamma-ray emission
is produced by the internal shock (S1) which is generated by the
collision of shell 1 and shell 2, while the prompt optical emission
can be produced by the internal shock (S2) which is generated by the
collision of shell 3 and shell 4, in this case the optical emission
and the gamma-ray emission is not correlated. In this paper we will
show that for GRB990123, GRB060111b and GRB041219a, the observed
optical flashes can all be explained by the internal shock model. We
also discuss the synchrotron-self-Compton emission in the internal
shock model.

\section{The emission from the internal shock}
In the standard fireball model of GRBs, the prompt gamma-ray
emission is produced by the internal shock, and the afterglow is
produced by the external shock. The internal shock model has been
discussed by many authors (e.g. Paczy\'nski \& Xu 1994; Daigne \&
Mochkovitch 1998). In the internal shock model, it is assumed that
the central engine emits lots of shells with different Lorentz
factors, the fast shell can catch up with the early slow shell and
then produce the internal shock.

\subsection{The synchrotron radiation of internal shock model}
If the typical Lorentz factors of the fast and slow shells are
$\Gamma_f$ and $\Gamma_s$ respectively, and the fast and slow shells
contain about the same masses, then the fast shell will catch up
with the slow shell at a radius $R\sim 2\Gamma^2 c\delta t/(1+z)$,
where $\delta t$ is the observed typical variability timescale. The
Lorentz factor of the merged shell is $\Gamma \approx \sqrt{\Gamma_f
\Gamma_s}$, and the Lorentz factor of the internal shock can be
estimated as $\Gamma_{sh} \approx
(\sqrt{\Gamma_f/\Gamma_s}+\sqrt{\Gamma_s/\Gamma_f})/2$ (Piran 1999).
Then the thermal energy density of the shocked material is $e
\approx 4\Gamma_{sh}(\Gamma_{sh}-1)n_e m_p c^2$ (Blandford \& Mckee
1976), where the comoving number density $n_e \simeq L_m/(4\pi
\Gamma^2 R^2 m_p c^3)$, $L_m$ is the outflow luminosity and $m_p$ is
the rest mass of proton. The strength of magnetic field is $B \simeq
9.3 ~{\rm G}~
(\frac{\epsilon_{B}}{0.5})^{1/2}[\Gamma_{sh}(\Gamma_{sh}-1)/2]^{1/2}
L_{m,52}^{1/2}(\frac{\Gamma}{600})^{-3}\delta
t_{1}^{-1}(\frac{1+z}{2})$ (Fan \& Wei 2005), where $\epsilon_B$ is
the energy fraction occupied by the magnetic field. Here the
convention $Q_x=Q/10^x$ has been adopted in cgs units throughout the
text.

As usual, the electrons accelerated by the internal shock would
follow the power law distribution $dn_e/d\gamma_e \propto
\gamma_e^{-p}$ for $\gamma_e > \gamma_{e,m}$, where
$\gamma_{e,m}=\epsilon_e (\Gamma_{sh}-1)[(p-2)m_p]/[(p-1)m_e]$ is
the minimum Lorentz factor of shocked electrons (Sari et al. 1998),
where $\epsilon_e$ is the energy fraction occupied by electrons,
$m_e$ is the rest mass of electron. Here we take $p=2.5$. Then the
observed typical frequency of the synchrotron radiation is (Fan \&
Wei 2005)

\begin{equation}
\nu_{m}\approx 3.2\times 10^{15} (\frac{\epsilon_e}{0.5})^2
(\frac{\epsilon_{B}}{0.5})^{1/2}
(\Gamma_{sh}-1)^{5/2}(\frac{\Gamma_{sh}}{2})^{1/2}
L_{m,52}^{1/2}(\frac{\Gamma}{600})^{-2}\delta t_{1}^{-1} ~{\rm Hz}
\end{equation}

The cooling Lorentz factor is $\gamma_{e,c}\simeq 7.7\times 10^{8}
(1+z)/[(1+Y)\Gamma B^2 \delta t]$ (Sari et al. 1998), where
$Y=[-1+\sqrt{1+4x\epsilon_e/\epsilon_B}]/2$ is the Compton
parameter, $x\simeq min\{1, (\nu_m/\nu_c)^{(p-2)/2}\}$ (Sari \& Esin
2001). Then the cooling frequency is

\begin{equation}
\nu_{c}\approx 5.7\times 10^{15}(\frac{1+z}{2})^{-2}
(\frac{\epsilon_{B}}{0.5})^{-3/2}[\Gamma_{sh}(\Gamma_{sh}-1)/2]^{-3/2}
L_{m,52}^{-3/2} (\frac{\Gamma}{600})^8 \delta t_{1} (1+Y)^{-2} ~{\rm
Hz}
\end{equation}

The synchrotron-self-absorption frequency is about (Li \& Song 2004)

\begin{equation}
\nu_{a}\approx 2.4\times 10^{14} (\frac{1+z}{2})^{-2/7}
(\frac{\epsilon_{B}}{0.5})^{1/14}
[\Gamma_{sh}(\Gamma_{sh}-1)/2]^{1/14}
L_{m,52}^{1/14}L_{syn,52}^{2/7} (\frac{\Gamma}{600})^{-8/7}\delta
t_{1}^{-5/7} ~{\rm Hz}
\end{equation}
where $L_{syn}$ is the synchrotron radiation luminosity. The maximum
flux of synchrotron radiation is $F_{max}\approx 3\sqrt{3}\Phi_p
(1+z)N_e m_e c^2 \sigma_T \Gamma B/(32\pi^2 q_e D_L^2)$, where $q_e$
is the charge of electron, $N_e=L_m \delta t/[(1+z)\Gamma m_p c^2]$
is the total number of emitting electrons, $\Phi_P$ is a function of
$p$, for $p=2.5$, $\Phi_P =0.6$ (Wijers \& Galama 1999). $D_L$ is
the luminosity distance, here we adopt
$(\Omega_M,\Omega_\Lambda,h)=(0.3,0.7,0.71)$. Using these equations,
we can discuss the synchrotron radiation features of GRBs.

\subsection{Application to GRB990123, GRB060111b and GRB041219a}
Recently we have shown that the optical flash and high energy
emission of GRB050904 can be explained by the emission of internal
shock (Wei et al. 2006). In this subsection we will discuss whether
the synchrotron radiation of internal shock can account for the
optical and gamma-ray emission of GRB990123, GRB060111b and
GRB041219a.

{\bf GRB990123} This is a very strong burst, the isotropic energy of
gamma-ray emission was about $3\times 10^{54}$ ergs (Andersen et al.
1999; Kulkarni et al. 1999). A well known feature of this GRB is
that a very bright optical flash was detected during the prompt
gamma-ray emission phase, this is the first time that a prompt
emission in another wavelength apart from gamma-rays has been
detected from a GRB (Akerlof et al. 1999; Kulkarni et al. 1999). The
slope of the power law decay of the optical flash is about -2 up to
10 minutes after the burst, then the flux decayed as $f\propto
t^{-1.1}$ which can be ascribed by the emission of external forward
shock. The optical flash was not correlated with the gamma-ray
emission, which implies they should arise from different regions.
Sari \& Piran (1999) have shown that the reverse shock emission can
explain the optical flash very well. However M\'esz\'aros \& Rees
(1999) have shown that the internal shock model can also explain the
temporal behavior of the optical flash. In their model (iii), the
outflow was assumed to be magnetic dominated, then the observed
optical flux would decay with time as $f_{\nu_{opt}} \propto t^{-p}$
if $\nu_m<\nu_{opt}<\nu_c$ (M\'esz\'aros \& Rees 1999). So if $p\sim
2$ we can get $f_{\nu_{opt}} \propto t^{-2}$, which is agreement
with the observation. However in that paper they only discussed the
scaling laws of the temporal behavior, while we need to know whether
the prompt optical emission can really be produced by synchrotron
radiation of the internal shock.

Using the equations given in the previous section, we find that if
we take the parameters as follows: $\Gamma\sim 800$, $L_{m,52}\sim
1$, $\delta t_{1}\sim 2$, $\epsilon_e \sim 0.3$, $\epsilon_B\sim
0.5$, then we get $\nu_m \sim 4\times 10^{14}$ Hz, $\nu_a \sim
5\times 10^{13}$ Hz, $\nu_c \sim 7\times 10^{16}$ Hz, $f_{\nu_m}\sim
8.5\times (\frac{1+z}{2})D_{L,28}^{-2}$ Jy, for $z=1.6$,
$f_{\nu_m}\sim 1$ Jy, which is quite agreement with the observation.
Since $\epsilon_B\sim 0.5$, the outflow is magnetic dominated, and
$\nu_m<\nu_{opt}<\nu_c$, so the relation $f_{\nu_{opt}} \propto
t^{-p}\propto t^{-2}$ is valid, which is consistent with the
observation.

After being accelerated by the shock, all the electrons will cool by
adiabatic expansion, so both the $\nu_m$ and $\nu_c$ will decrease
with time as $t^{-2}$ (M\'esz\'aros \& Rees 1999), then there is a
question when will the cooling frequency $\nu_c$ cross the optical
band? Because we know that if $\nu_c$ has crossed the optical band
then the optical flux will drop sharply. The optical flash was
occurred at about 50 seconds after the burst, and its emission
lasted to about 600 seconds. So at 50s the cooling frequency is
$\nu_c \sim 7\times 10^{16}$ Hz, then at 600s $\nu_c \sim 5\times
10^{14}$ Hz, which is still larger than the optical band. Therefore
we conclude that the bright optical flash of GRB990123 can be well
explained by the internal shock model.

{\bf GRB060111b} This was a bright, double-peak gamma-ray burst with
duration about 60 seconds, the fluence in the 15-350KeV band was
$1.6\times 10^{-6}$ ergcm$^{-2}$. Very recently Klotz et al. (2006)
presented the early optical emission of this GRB, which is the first
time that the early optical emission was monitored with a temporal
resolution of a few seconds during the prompt high energy emission
phase. They reported that from 28s to 80s after the trigger, the
optical flux decayed with a slope $\sim -2.38$, then it was followed
by a shallow decay with index $\sim -1.08$, but if we assume that
the optical emission was the superposition of two components, then
the slopes of the fast and slow decay become -3 and -0.9
respectively (Klotz et al. 2006). These features are very similar to
the case of GRB990123, so we believe the early optical emission of
GRB060111b can also be explained by the internal shock model.

At 28s after the trigger, the observed optical emission was about
13.75 magnitude, however, Klotz et al. (2006) pointed out that an
extinction of $A_R=4$ magnitudes was required to reconcile the
measured optical flux with the XRT spectrum, so the intrinsic
optical flux at 28s was about 0.5 Jy. Since the redshift of
GRB060111b is not available, we take $z=1$. Considering the
similarity between this burst and GRB990123, we adopt the same
parameters as GRB990123 except $L_m$, now we take $L_{m,52}=0.5$,
then we obtain $\nu_m \sim 2.8\times 10^{14}$ Hz, $\nu_a \sim
4.8\times 10^{13}$ Hz, $\nu_c \sim 2\times 10^{17}$ Hz,
$f_{\nu_{opt}}\sim 2.3 (\frac{1+z}{2})D_{L,28}^{-2}$ Jy, for $z=1$,
$f_{\nu_{opt}}\sim 0.5$ Jy, which is quite consistent with the
observation.

According to the model (ii) and (iii) of M\'esz\'aros \& Rees
(1999), in the case of $\nu_m<\nu_{opt}<\nu_c$, the flux could decay
with time as $F_{\nu}\propto t^{(1-3p)/2}$ or $F_{\nu}\propto
t^{-p}$, so for the reasonable values of $p$, the observed optical
emission can be well explained by the internal shock model.

{\bf GRB041219a} This GRB was detected by both the INTEGRAL and
Swift satellites. The 15-350KeV fluence was about $1.55\times
10^{-4}$ ergs cm$^{-2}$, placing it among the top few percent of the
whole GRB catalog (Vestrand et al. 2005; Blake et al. 2005). It was
also one of the longest GRBs with duration about 520 seconds.
Between 200-400 seconds one optical flash temporally coincident with
the gamma-ray emission was detected. After correcting for the
nominal extinction, the peak optical magnitude is about $R_c \sim
13.7$. In contrast to GRB990123, the optical flash of GRB041219a
seems to be correlated with the gamma-ray emission, which strongly
suggests that they should have the same origin.

One possibility is that the optical flash was produced by the
synchrotron radiation of the internal shock, while the gamma-ray
emission was produced by the synchrotron-self-Compton process, just
as the case of GRB050904 (Wei et al. 2006). However the observation
shows that the gamma-ray fluence is much larger than that of the
optical emission, $F_{\gamma}/F_{opt}\sim 10^{5}$, so if the
gamma-ray emission was produced by the SSC process, then the Compton
parameter $Y$ should be about $\sim 10^{5}$, which is unreasonable
large. Therefore it is more likely that the prompt optical emission
was the low energy tail of the high energy emission.

The spectral analysis shows that in the range 15-350KeV the slope of
the spectra is about -1/2, which implies that $\nu_c$ or $\nu_m$
should be larger than 350KeV. We find that if $\nu_m>350$KeV, then
the self-absorption frequency $\nu_a$ is also very large, then the
optical flux would be too low. So the only possible case is $\nu_c
>350$ KeV. If we take the parameters as follows: $\Gamma\sim 1000$, $L_{m,52}\sim
0.1$, $\delta t_{1}\sim 0.05$, $\epsilon_e \sim 0.3$,
$\epsilon_B\sim 0.02$, then we get $\nu_m \sim 3\times 10^{14}$ Hz,
$\nu_a \sim 10^{15}$ Hz, $\nu_c \sim 7\times 10^{19}$ Hz,
$f_{\nu_m}\sim 27.5\times (\frac{1+z}{2})D_{L,28}^{-2}$ mJy. The
redshift of GRB041219a is not available, but we find that, in order
to account for the observation, its redshift cannot be large. If we
take $z\sim 0.2$, then $f_{\nu_m}\sim 275$ mJy, $f(200KeV)\sim 1$
mJy, $f_{\nu_{opt}}=f_m (\frac{\nu_a}{\nu_m})^{-1/2}
(\frac{\nu_{opt}}{\nu_a})^{5/2}\sim 10$ mJy, these results are
agreement with the observation. We note that both the fluxes in the
gamma-rays and optical and the spectral slope are all consistent
with the observation. It is interesting to note that Barkov \&
Bisnovatyi-Kogan (2005) restricted the redshift of GRB041219a as
$z\sim 0.12$ by fitting the observed IR afterglow. Very recently
McBreen et al. (2006) estimated the pseudo-redshift of this burst
and found the lower limit is $z\sim 0.3$.

\section{The synchrotron-self-Compton emission of the internal shock}

Up to now we only consider the synchrotron radiation in the internal
shock, one may ask whether the synchrotron-self-Compton (SSC)
process would play an important role in the internal shock model?
We'll give a brief discussion on this topic.

If the internal shock is in fast cooling phase, i.e. $\nu_c<\nu_m$,
and if $\nu_m$ is around the optical band, then the IC scattered
photons will peak at $\nu_{m,IC}\sim 2\gamma_{e,m}^{2}\nu_m \sim
100$ KeV, and the peak flux of SSC emission is $f(\nu_{m,IC})\sim
Y\nu_m f(\nu_m)/\nu_{m,IC} \sim Yf(\nu_m)/\gamma_{e,m}^{2}\sim
10^{-4}f(\nu_m)$ ($Y$ is usually of order unity). We note that at
$\nu_{m,IC}$ the SSC emission will exceed the synchrotron emission
when $p>2$. If $f(\nu_m)\sim 50$ mJy, then the IC flux at 1 KeV
would be $\sim 0.05$ mJy, which is just the case of GRB050904
(Bo\"{e}r et al. 2006; Wei et al. 2006). If $\nu_m\sim 1$ KeV, i.e.
the synchrotron radiation produced the observed X-ray flare, then
the IC scattered photons will peak at $\nu_{m,IC}\sim
2\gamma_{e,m}^{2}\nu_m \sim 10$ MeV (see also Wang et al. 2006), the
IC peak flux is still $f(\nu_{m,IC})\sim Y\nu_m f(\nu_m)/\nu_{m,IC}
\sim 10^{-4}f(\nu_m)$, then at 100 KeV the IC flux is about
$10^{-3}f(\nu_m)$. However at 100 KeV the synchrotron emission flux
is $\sim 10^{-p}f(\nu_m)$, so if $p<3$ (which is usual the case)
then at 100 KeV the IC flux is lower than the synchrotron radiation.
The SSC flux will be dominant when $\nu>10$ MeV. For GRB011121, Piro
et al. (2005) reported the detection of a X-ray flare with flux
$\sim 1$ mJy, then we predict there should be a corresponding flare
at about 10 MeV with fluence $\sim 10^{-9}$ erg cm$^{-2}$ s$^{-1}$.

On the other hand, if the internal shock is in slow cooling phase,
then from eqs.(1)(2) we find that the Lorentz factor $\Gamma$ would
be very large, for typical parameters $\Gamma$ should be larger than
500, for example, for GRB990123 we obtain $\Gamma\sim 800$. In this
case, the cooling Lorentz factor $\gamma_{e,c}$ is also very large,
$\gamma_{e,c}\sim 10^{3}$. For the case of GRB990123, $\nu_m \sim
10^{14}$ Hz, $\nu_c \sim 5\times 10^{16}$ Hz, then the peak
frequency of SSC is $\nu_{c,IC}\sim 2\gamma_{e,c}^{2}\nu_c \sim 1$
GeV, the IC peak flux is $f(\nu_{c,IC})\sim Y\nu_c
f(\nu_c)/\nu_{c,IC} \sim 10^{-6}f(\nu_c)$, while at $\nu_{c,IC}$ the
flux of synchrotron emission is $\sim 10^{-3p}f(\nu_c)$, so if
$p\sim 2$, then the SSC component is always unimportant. If $p\sim
2.5$, then the SSC component will dominate the synchrotron radiation
when $\nu >1$ MeV. Therefore for the case like GRB990123, we expect
there would be a GeV flare with fluence about $10^{-8}$ erg
cm$^{-2}$ s$^{-1}$, which might be detected by the GLAST satellite.

\section{Discussion and conclusion}
The origin of the prompt optical emission contemporaneous with the
high energy emission is a very important issue, but it is still
unclear. Since the information about the GRB central engine has been
lost in the later afterglow, so it is crucial to study the early
afterglow and prompt emission. However, for these GRBs with prompt
optical emission detected, the relationship between the optical and
gamma-ray flux is quite different. For GRB990123and GRB060111b, the
optical and gamma-ray emission vary independently, and the optical
flux is much higher than the back extrapolation of the late
afterglow. For GRB041219a and GRB050904, the optical and high energy
emission are correlated, but for GRB041219a it is rather unlikely
that the gamma-ray emission is produced by the SSC process, while
for GRB050904 the high energy emission can be attributed to the SSC
emission (Wei et al. 2006). GRB050401 is another GRB with prompt
optical emission detected, its optical emission was uncorrelated
with the gamma-ray emission, the most unique feature is that its
optical emission can be well fit by the back extrapolation of the
late afterglow emission, which suggests that the prompt optical
emission may be from the external forward shock emission (Rykoff et
al. 2005), so in this paper we did not discuss this burst.

The early optical flash of GRB990123 is widely believed to be
produced by the reverse shock emission(Sari \& Piran 1999), and it
is suggested that the reverse shock is magnetized (Fan et al. 2002;
Zhang et al. 2003). The reverse shock model has also been used to
explain some other GRBs (Wei 2003; Zhang et al. 2003; Fan et al.
2005; Shao et al. 2005). If the reverse shock emission is strong
like GRB990123, then it is expected that the prompt optical flash
would be easily detected by Swift. However, the observations show
that for a large fraction of the GRBs no early optical emission has
been detected, which implies that the reverse shock emission should
be significantly suppressed (Roming et al. 2005). One possibility is
that the GRB outflow may be Poynting flux-dominated (Fan et al.
2004; Zhang \& Kobayashi 2005).

A different origin for the simultaneous optical flash is that the
optical flash may arise from the internal shock emission
(M\'esz\'aros \& Rees 1999; Fan \& Wei 2004). Between the time at
which the internal shock stops and the time when the outflow is
decelerated by the surrounding medium, the bulk Lorentz factor
remains approximate constant, hence under some circumstances the
emission during this period can account for the observed optical
flash. In particular, for a more realistic situation, the average
Lorentz factor of the shells could vary. For the case that average
$\Gamma$ increases with time, then a power-law decay of the optical
light curve with index $\sim 2$ can be obtained even for $p>2$.

It should be noted that the internal shock model is always used to
explain the prompt gamma-ray emission, however we note that in the
internal shock model, the typical synchrotron radiation frequency
strongly depends on the parameters, such as $\Gamma$, $\Gamma_{sh}$,
$L_m$, $\delta t$ etc., and for different shocks it is natural that
these parameters are different, so we expect that the internal shock
model not only can produce the gamma-ray emission, but also can
produce the optical or X-ray emission. One good example is that the
bright X-ray flares observed in nearly half of the GRBs has been
well explained by the late internal shock model (Fan \& Wei 2005;
Zhang et al. 2006).

Here we discussed the emission features of optical flash based on
the standard internal shock model, and found that the observations
can be well explained by the internal shock model. We found that for
these GRBs with optical flash, the values of $\delta t$ and $\Gamma$
are usually larger than that of the typical GRBs. For typical GRBs
$\delta t \sim 10^{-2}$ s and $\Gamma \sim 300$. We suggest this may
explain why the optical flashes like GRB990123 have not been
observed for a number of GRBs, this is because in order to produce
the optical flash, the time interval between the two shells should
be large, while in this case the shells would be easily disrupted by
other shells, so only those survived shells can collide to produce
the optical flash. If this is true, then we expect the bright
optical flash like GRB990123 should not be common in GRBs.

In the internal shock model, the value of $\delta t$ should be
determined by the physical processes which powered the gamma-ray
burst. However up to now the explosion mechanism is still unclear,
so the value of $\delta t$ cannot be obtained theoretically. But
fortunately, the value of $\delta t$ can be estimated from the
observed light curve, since $\delta t$ reflects the typical
variability timescale of the light curve. From observations we note
that the variability timescale of optical flash is much longer than
that of gamma-ray emission, so it is quite reasonable that $\delta
t$ of optical flash is much longer than that of gamma-ray emission.

We also discuss the synchrotron-self-Compton process in the internal
shock model, we have shown that there are several possibilities
depending on the values of parameters. In the fast cooling case, if
$\nu_m$ is around the optical band, then there would be a flare
occurred at $\sim 100$ KeV. If $\nu_m \sim 1$ KeV, i.e. the X-ray
flare observed in many GRBs, then the SSC emission will be dominant
at the frequency $\nu\sim 10$ MeV. While in the slowing cooling
case, since $\Gamma$ is very large, there would be a GeV flare
accompanying the optical flash, and for GRB990123 a GeV flare with
fluence about $10^{-8}$ erg cm$^{-2}$ s$^{-1}$ is expected, which
might be detected by the GLAST satellite. The more details on the
synchrotron-self-Compoton emission accompanying the optical flashes
needs numerical calculation, and we will give a detailed numerical
calculation in a subsequent paper.

\section{acknowledgements}
I thank the referee for her/his helpful comments. I thank Y.Z. Fan
for helpful discussions. This work is supported by the National
Natural Science Foundation (grants 10225314 and 10233010) of China,
and the National 973 Project on Fundamental Researches of China
(NKBRSF G19990754).

\end{document}